\documentclass[10pt]{article}
\usepackage{balance}

\usepackage{geometry}
\usepackage{hyperref} 
	\hypersetup{
       		 citecolor=black,
		pdfstartview=true,
		colorlinks=true,
		urlcolor=blue            
	}
\usepackage{amsthm}
\usepackage{enumerate}
   
\begin{document}
 
\title{A Framework for Blockchain-Based Applications}

\author{Ephraim Feig\\
IEEE Life Fellow\\
\date{}
}

\maketitle

\thispagestyle{plain}
\pagestyle{plain}

\begin{abstract}
Blockchains have recently generated explosive interest from both academia and industry, with many proposed applications. But descriptions of many these proposals are more visionary projections than realizable proposals, and even basic definitions are often missing. We define ``blockchain'' and ``blockchain network'', and then discuss two very different, well known classes of blockchain networks: cryptocurrencies and Git repositories. We identify common primitive elements of both and use them to construct a framework for explicitly articulating what characterizes blockchain networks. The framework consists of a set of questions that every blockchain initiative should address at the very outset. It is intended to help one decide whether or not blockchain is an appropriate approach to a particular application, and if it is, to assist in its initial design stage.
\end{abstract}

\section{Introduction}\label{introduction}   
Blockchain has been heralded as ``a foundational technology: It has the potential to create new foundations for our economic and social systems''\cite{hbr}. People claim that blockchain will streamline the electronic health records process \cite{wired_ehr}; will ``add greater visibility and efficiency across the entire supply chain to deliver higher value to your customers and trading relationships''\cite{ibm_sc}; will track ownership of real estate \cite{fastcompanyre}; may ``disrupt the insurance industry and change the way we share data, process claims and prevent fraud''\cite{forbes_insurance}; ``could revolutionize the Internet of Things'' \cite{forbes_iot}.
 
There are no explicit description of the blockchains in the cited applications. But the blockchains of cryptocurrencies are well understood. As Satoshi Nakamoto writes \cite{satoshi}, they are needed to enable ``electronic transactions without relying on trust.'' A complete, immutable public record of transactions is not a design goal in cryptocurrencies. Quite the contrary, cryptocurrency purists would prefer no record at all, to strengthen identity-hiding. Nakamoto wrote that, as was well known already  \cite{dai}, ``To accomplish this without a trusted party, transactions must be publicly announced.''  The public ledger is the price to pay in order to enable cryptocurrency with integrity.  
 
What exactly is the trust that Nakamoto was referring to in his paper? In an online forum \cite{satoshi2} he writes,``The root problem with conventional currency is all the trust that's required to make it work. The central bank must be trusted not to debase the currency, but the history of fiat currencies is full of breaches of that trust. Banks must be trusted to hold our money and transfer it electronically, but they lend it out in waves of credit bubbles with barely a fraction in reserve. We have to trust them with our privacy, trust them not to let identity thieves drain our accounts."
 
The central bank and fiat currencies issues that Nakamoto bemoans pale in comparison to their analogues in the cryptocurrency world. Anybody can create a cryptocurrency (there are over 1,500 of them \cite{coinmarketcap}) and even the most established ones will fork at the whims of their creators \cite{forks}. As for banks' abuses, relative to total currency volume, they pale in comparison to Ponzi schemes by cryptocurrency exchanges \cite{ponzi}. As for draining accounts, cryptocurrencies lose by a wide margin \cite{hacks}. The concern that we have to trust banks to hold and transfer funds is genuine, but mostly in cases where the transfer is prohibited by some legal authority. The big advantage cryptocurrencies offer is identity-hiding, and only to those who are able to cover their tracks \cite{criminals}. To gain this advantage, Bitcoin needed a blockchain with proof-of-work (PoW). The cost for this advantage today (3/1/2018) is estimated at 791 KWh of electricity consumed per transaction \cite{energy}, slow batch-processing of transactions, plus the loss of such standard payment services as fraud protection and loss recovery.

The enormous costs of operating the Bitcoin network are recognized, and new types cryptocurrencies are being deployed to deal with them. Those that use proof-of-stake (PoS) instead of PoW are emerging \cite{PoS_coins}, and Ethereum is expected to implement its version \cite{Buterin} in the near future. PoS adds an escrow component to the network \cite{escrow}, bringing it a little closer to standard payment systems, while still bypassing central authority. Blockchains for applications other than currencies go even further in conforming to standard trust models. Permissioned-blockchains have been suggested for enterprise applications, in which ``participants need to obtain an invitation or permission to join. The access control mechanism could vary: existing participants could decide future entrants; a regulatory authority could issue licenses for participation; or a consortium could make the decisions instead."\cite{IBM_private} Moving even further from trust averse systems, we have private networks that are not only permissioned but also restrict who can see the blockchain.

Git may be the most successful blockchain platform \cite{github_stats} and has been operational over a decade. It is the most popular version control system. While it is used primarily for team development of software, it can be used to maintain any set of text-based records. The software that powers Bitcoin is developed on Git \cite{github_bitcoin} as an open source project. Every branch of a Git repository is a blockchain with commits as blocks. Peers are software developers, and they all should maintain the master branch of the repository. Git is permission-based, and consensus is based on trust; there is no mining or proof-of-anything. Git blockchains of open-source projects are visible to the public; private repositories restrict who can see them.
 
Unlike Bitcoin, where forking is an inadvertent, uncommon event, with Git, branching is modus operandi. Unlike Bitcoin, where forks are abandoned upon acceptance of a longest blockchain, in Git, branches are frequently merged. And strikingly, unlike Bitcoin, where there is total aversion to central authority, even though Git is a peer-to-peer system, Git projects are almost always used in conjunction with some Git hosting service such as  GitHub, Bitbucket, Perforce or CloudForge \cite{services}. 

People disagree whether or not Git branches are blockchains \cite{git_bc1, git_bc2, git_bc3, git_bc4} because there have been only few definitions proposed in this new field, and these have not been consistent. We like the approach in \cite{game}, where the term ``blockchain'' refers to a data-structure, and the term "blockchain network" refers to how that data structure is deployed. We also want the definitions sufficiently broad so as to include cryptocurrencies and Git branches, which operate in totally different trust environments.  

\textbf{Definition:}
A \emph{blockchain} is a sequence of blocks of data in which each block, other than the first, is cryptographically linked to its predecessor.  
 
Cryptographic linking is well understood \cite{linking}. We are not specifying how the blocks are cryptographically linked. Bitcoin linking is described in \cite{btc_link} and Git linking in \cite{git_link}.  

\textbf{Definition:}
A \emph{blockchain network} is a peer-to-peer network in which peers collaborate to achieve a common goal by using a blockchain.  
 
The Bitcoin system is a blockchain network. A Git repository is a blockchain network, with the master branch as the blockchain used in collaboration. While Git peers often fork the blockchain by creating new branches, they collaborate using the master branch and strive to merge their other branches with it. In general, we do not insist that peers all maintain identical or complete blockchains, or that the network maintain an immutable blockchain, though these may be the common goal in some applications.

\section{Background}
Bitcoin and Git blockchains have a fundamental data structure in common- blocks connected by cryptographic hashes- and a  fundamental difference- the presence or absence of proof-of-work (PoW). Both cryptographic hashing and PoW have been studied a long time before the word blockchain was introduced. It is instructive to see how they were deployed and then ask, why all of a sudden, the two of them combined are generating so much enthusiasm?  
 
The idea of chaining blocks of data together with cryptographic hashes has been around since the late 1970's. By 1982, when Ralph Merkle's patent \cite{merkle} was granted, cryptographic protocols were already evolving \cite{merkle2}. The data structure named after him, the Merkle Tree, found utility in peer-to-peer systems in which peers all needed to share identical data \cite{merkle3,merkle4}. Any change in the data would be very quickly detected. In most cases, when a change is detected, there is recourse, the change can be undone. Peers who want the change can accept it; peers who do not want the change, can undo it. If peers cannot reach consensus, they can diverge along distinct paths. 
 
The Bitcoin blockchain is a binary Merkle Tree in which every right branch contains only one leaf, considered as a sequence from left to right. Every Bitcoin block stores transaction data as a Merkle Tree. Therefore, any change is easily detectable. But a bitcoin transaction is irreversible, and there is no remedy to a double spend. Therefore, to preserve Bitcoin's integrity, Nakamoto had to introduce PoW. Fortuitously, PoW also provided the mechanism for disseminating new bitcoins and incentivizing peers. 
 
PoW was introduced by C. Dwork and M. Naor in 1992 \cite{dwork} to combat email spam. More generally, PoW can be used to deter people from doing undesirable things by making them very expensive. PoW was proposed to mitigate distributed denial-of-service attacks \cite{denial}. In 2003, PoW was proposed to provide incentives in peer-to-peer systems \cite{incentives}. The particular PoW used in Bitcoin is a variant of Adam Back's Hashcash \cite{hashcash}, proposed in 1997, which requires finding a nonce that yields a hash with some leading 0's.  
 
Until Bitcoin, PoW did not gain any significant traction. A study of its original proposed application, spam prevention, concluded that PoW ``does not work'' \cite{spam}. With Bitcoin, things changed. There were users who insisted on a payment system in which participants hide their identities, and for them, the gains outweighed the costs.

 \section{Blockchain Network Dynamics}\label{dynamics}
 
The blockchain networks of cryptocurrencies and Git are characterized by the following activities:   

\setlength{\parskip}{0em}

\begin{enumerate}
	\item Developers create and deploy the network.
\end{enumerate}
Once the network is deployed,
\begin{enumerate}
\setcounter{enumi}{1}
	\item Users input data into the network
	\item Peers propose blocks to add to the blockchain 
	\item Peers validate proposed blocks
	\item Peers strive to reach consensus
\end{enumerate}
They are also different in two fundamental ways:
\begin{enumerate}
\setcounter{enumi}{5}
	\item Whether or not there are irreversible inputs
	\item Whether or not the blockchain is immutable
\end{enumerate}
And in all cases,
\begin{enumerate}
\setcounter{enumi}{7}
	\item Peers are incentivized to perform
\end{enumerate}

\setlength{\parskip}{0.6em}

We see three types of participants: developers, users and peers. While developers play a critical role, they are not part of the framework. The framework relates to the steady-state network once it is operational.  

With cryptocurrencies, peers are users but most users are not peers. Most users just input transactions. When they receive payments, they have no interaction with the blockchain. The blockchain being public, everybody can be a user by just looking at it. However, one cannot use the blockchain to prove payment to a particular payee because one cannot force a payee to provide proof of received payment. There is a whole class of users who analyze cryptocurrency blockchains to catch illegal behavior such as money laundering and tax evasion. These include US departments DEA, FBI, ICE, IRS and SEC \cite{forensics}. 
 
In Git, users who enter data into the blockchain are also peers. They input snippets of code and together they create and maintain some desired code. Git users who are peers do not hide their identities, but quite the opposite, seek recognition for their contributions. Another set of users take the code and deploy it. 

In Bitcoin, after a peer succeeds a PoW, it will propose a block comprising a set of transactions, the size limited by the Bitcoin protocol. The Bitcoin incentives are such that the peer will validate the block before proposing it to the network. Almost always, even though all peers are working hard on their proofs, only one gets to propose a block (in the rare case of a fork, two or more peers will each propose a block) and the other peers will validate it. Immediately afterwards, the peers start all over again, work on their proofs, and when a peer succeeds, it proposes a new block of transactions. One can see the enormous costs here, as at every cycle, all the peers' duplicate efforts are essentially wasteful but necessary to maintain the integrity of the system. 

In Git, a block is a commit, with data describing changes made to the code. These changes take the form of additions to and deletions from the code. There are no size limitations to a commit. Several peers may simultaneously submit blocks of code that they have been working on individually, and all of these may be added to the blockchain. 

In Bitcoin, validation involves a fixed, deterministic set of rules \cite{validate1} and is accomplished automatically, without human intervention. Peers validate that payers have sufficient funds to cover their payments, and to do so, they have to verify that the bitcoins used for payment are still in circulation. As Nakamoto pointed out \cite{satoshi}, ``The only way to confirm the absence of a transaction is to be aware of all transactions.''  This is why a Bitcoin peer needs the entire blockchain.

In Git, validation is more nuanced and involves human activity. There are no fixed, deterministic rules that specify validity for a block. When a new block is proposed to add to the master branch, peers must check that it does not crash the code. In order to do that, they must have the entire code. This is why a Git peer needs the entire master branch. Peers cannot guarantee that a block does not crash the code; they can only say they tested it sufficiently and are comfortable with proceeding forward; if a bug is later found, somebody will probably fix it. They may also validate that the proposed code actually does what the user wrote in the message part of the commit that it should do, and that what it does is desirable. 

Consensus in Bitcoin is automatic. Peers choose the candidate blockchain of greatest height. The Bitcoin protocol with its PoW guarantees that, as long as peers possessing a majority of the hash power in the network validate truthfully, then asymptotically, the blockchain with greatest height will be valid. Note that it is a majority of hash power in the network, not a majority of peers, that is necessary for users to trust the Bitcoin network. If there is no consensus, the blockchain may fork and peers may continue with either or both branches of the fork.  

Consensus in Git is not automatic. Peers are structured with hierarchical authority. They may know each other, converse with each other, and often defer to peers with the most significant contributions. Git uses an interesting form of PoS; the stake includes a peer's reputation, permission level, and membership in the development team. As with Bitcoin, if no consensus is reached, the project may fork and peers may continue as they choose.

Bitcoin transactions are irreversible. Once a payment is made, there is no recourse. With Git, if a user commits a block with a subtle bug that the peers did not catch, somebody can fix that bug in a later commit. This is not an uncommon Git event. 

The Bitcoin blockchain is practically immutable. With Git, by design, one can rewrite history \cite{git_hist}. 

Cryptocurrency peers are incentivized with mining rewards and transaction fees. Git peers on open-source projects are rewarded by getting the code they desire and by recognition. On closed projects, peers are often salaried software developers.

 \section{The Framework}\label{framework}
 
The framework is a set of questions that ask for specifics relating to blockchain network dynamics. It should be used at the conceptual stage, before development, of a blockchain-based application.
\begin{center}
\fbox{ 
\parbox{8cm} {

\begin{enumerate}
\item Who are the users?
\item What data do users input?
\item Are any inputs irreversible?
\item Who are the peers?
\item How do peers create blocks?
\item What do peers validate?
\item How do peers validate?
\item How do peers reach consensus?
\item Is the blockchain immutable?
\item How are peers incentivized?
\end{enumerate}
}
}
\end{center}

\textbf{Who are the users?} Do users have a task that current systems not handle well? Do they want to do something that they could not have done without a blockchain? Are they primarily concerned that their data is at risk? Are they anticipating reduced costs? What performance expectations, like transaction latency, do they expect? Will users have access restrictions to the blockchain? Will there be one blockchain for several classes of users or several blockchains, each for one class of users?

\textbf{What data do users input?} Do users enter blobs (large data sets) or transactional data with pointers to blobs that are stored elsewhere? Will there be one blockchain for several types of transactions or several blockchains, each for a particular type of transaction?  
  
\textbf{Are any inputs irreversible?} For these, what are the potential liablilities? 
 
\textbf{Who are the peers?} Who can become a peer? How many peers are necessary? What is a desirable number of peers? Who are the initial peers? How much trust is there amongst the peers? Will peers have access restrictions to the blockchain? Are some peers also peers in other blockchain networks?

\textbf{How do peers create blocks?} How many transactions or contracts or whatever go into a block? When are blocks formed? Do blocks have size limitations? Is the data in the blocks structured or unstructured or both? Do peers have to process data before putting it in a block? 

\textbf{What do peers validate?} Do peers validate that the data satisfies certain conditions aside from syntactical constraints and/or proof-of-something? Is validation deterministic or do peers sometimes have to make judgement calls? 

\textbf{How do peers validate?} What information do peers need in order to validate? Do they need the blockchain to be able to validate? Do they need the entire blockchain? Can peers communicate with each other during validation?

\textbf{How do peers reach consensus?} What is the trust model? Can peers communicate with each other during the consensus process? 

\textbf{Is the blockchain immutable?} If so, what are the potential liabilities?

\textbf{How are peers incentivized?} Do peers compete for rewards? Do users compete to get peers to service their inputs?

\section{Discussion}
People have recently become enamored with avoiding reliance on trust. More accurately, they think they can create distributed networks of untrusting peers that are more trustworthy than individuals or organizations. They think that complete histories of records, even personal ones, should be on public display, with identities hidden. They want contractual or compliance obligations automatically validated without third party intervention. They see blockchain as the driving technology enabling all this.

As we have seen, the core technologies behind blockchain have been known for decades. Cryptographic linking of data is used in peer-to-peer file sharing such as BitTorrent \cite{bit_tor} and data and software distribution \cite{sw_dis}. They do not need proof-of-work because any tampering could be quickly detected and dealt with. Things changed with Bitcoin, where double spending, even if detected immediately, could not be undone. Bitcoin introduced a novel twist, combining cryptographic linking with proof-of-work, and after percolating for several years, the concept is now generating tremendous excitement. 

Git shows that blockchains can be successful in environments of trust. It makes sense, then, not to restrict the definition of blockchain to any particular trust model. Different trust models will lead to different validation and consensus protocols. Open source projects that are hosted on GitHub provide examples of trust-based consensus practices. For example, see the instructions to prospective peers relating to consensus in Bitcoin Core \cite{bitcoin_instruct}. Contrast that with the instruction to peers of the Git platform software \cite{git_instruct}.

When designing a business application, one should ask, where in the application do fraud or accidental errors happen? In many cases, they hardly ever happen to the data once it is stored, but typically happen during data entry. In such cases, validation may require some level of trust. For example, in supply-chain applications, if peers are not actually present when transactions occur (like boxes loaded on a truck), how will they validate that data entered are true records of what happened? 

Immutability, a requirement in almost all currently proposed blockchains, can be a liability. For example, with a public blockchain, if a secret key is compromised, then all immutable records associated with that key are permanently exposed. With standard systems, if a password is compromised, records associated with that password are exposed until the password is changed. Relaxing the requirement for immutability may be an appropriate choice for some blockchain-based applications.  

Irreversibility, which is considered an advantage in certain applications, can also be a liability. With Bitcoin, a payer cannot even use the blockchain to prove who is the payee. If a bitcoin owner loses a private key, the coin associated with it is permanently lost. There is no equivalent to a password-reset in Bitcoin. It is estimated that millions of bitcoins are already lost forever \cite{lost}. 

\section{Conclusion}
Blockchain-based applications other than cryptocurrencies and Git are at a very early stage, and there is clearly enormous enthusiasm to explore and create them. We foresee a spectrum of trust models evolving, and blockchain networks with validation and consensus protocols corresponding to them. We defined blockchains and blockchain networks, expanding the usual definition by not requiring immutability. We looked at two motivating examples with polar opposite trust models- Bitcoin and Git.  By considering their common properties, we created a framework to compel concreteness in discussing proposed blockchain-based applications and to guide in their early design stages.

\balance

\end{document}